# Ten Quick Tips for Deep Learning in Biology

*This manuscript ([permalink](#)) was automatically generated from [Benjamin-Lee/deep-rules@2d2f859](#) on May 27, 2021.*

## Authors

- **Benjamin D. Lee**
  [0000-0002-7133-8397](#) · [Benjamin-Lee](#)
  In-Q-Tel Labs; School of Engineering and Applied Sciences, Harvard University; Department of Genetics, Harvard Medical School

- **Anthony Gitter**
  [0000-0002-5324-9833](#) · [agitter](#) · [anthonygitter](#)
  Department of Biostatistics and Medical Informatics, University of Wisconsin-Madison, Madison, Wisconsin, USA; Morgridge Institute for Research, Madison, Wisconsin, USA · Funded by National Science Foundation (DBI 1553206); National Institutes of Health (R01GM135631)

- **Casey S. Greene**
  [0000-0001-8713-9213](#) · [cgreene](#)
  Department of Systems Pharmacology and Translational Therapeutics, Perelman School of Medicine, University of Pennsylvania, Philadelphia, PA, USA; Department of Biochemistry and Molecular Genetics, University of Colorado School of Medicine, Aurora, CO, USA; Center for Health AI, University of Colorado School of Medicine, Aurora, CO, USA · Funded by National Institutes of Health (R01 HG010067); the Gordon and Betty Moore Foundation (GBMF 4552)

- **Sebastian Raschka**
  [0000-0001-6989-4493](#) · [rasbt](#) · [rasbt](#)
  Department of Statistics, University of Wisconsin-Madison · Funded by Wisconsin Alumni Foundation (AAD5912)

- **Finlay Maguire**
  [0000-0002-1203-9514](#) · [fmaguire](#)
  Faculty of Computer Science, Dalhousie University · Funded by Donald Hill Family Fellowship

- **Alexander J. Titus**
  [0000-0002-0145-9564](#) · [AlexanderTitus](#) · [1alexandertitus](#)
  University of New Hampshire; Bioeconomy.XYZ

- **Michael D. Kessler**
  [0000-0003-1258-5221](#) · [mdkessler](#)
  Department of Oncology, Johns Hopkins University; Institute for Genome Sciences, University of Maryland School of Medicine · Funded by National Institutes of Health (R01DE027809)

- **Alexandra J. Lee**
  [0000-0002-0208-3730](#) · [ajlee21](#)
  Genomics and Computational Biology Graduate Program, University of Pennsylvania; Department of Systems Pharmacology and Translational Therapeutics, University of Pennsylvania · Funded by the Gordon and Betty Moore Foundation (GBMF 4552)

- **Marc G. Chevrette**
  [0000-0002-7209-0717](#) · [chevrm](#) · [wildtypeMC](#)
  Wisconsin Institute for Discovery and Department of Plant Pathology, University of Wisconsin-Madison · Funded by Grant 2020-67012-31772 (accession 1022881) from the USDA National Institute of Food and Agriculture

- **Paul Allen Stewart**
  0000-0003-0882-308X · pstew · biodataguy
  Department of Biostatistics and Bioinformatics, Moffitt Cancer Center, Tampa FL · Funded by Moffitt Cancer Center Support Grant (P30-CA076292)

- **Thiago Britto-Borges**
  0000-0002-6218-4429 · tbrittoborges
  Section of Bioinformatics and Systems Cardiology, Klaus Tschira Institute for Integrative Computational Cardiology, University Hospital Heidelberg; Department of Internal Medicine III (Cardiology, Angiology, and Pneumology), University Hospital Heidelberg

- **Evan M. Cofer**
  0000-0003-3877-0433 · evancofer · evan_cofer
  Lewis-Sigler Institute for Integrative Genomics, Princeton University, Princeton, NJ, USA; Graduate Program in Quantitative and Computational Biology, Princeton University, Princeton, NJ, USA · Funded by National Institutes of Health (T32 HG003284); National Science Foundation Graduate Research Fellowship Program

- **Kun-Hsing Yu**
  0000-0001-9892-8218 · khyu
  Department of Biomedical Informatics, Harvard Medical School; Department of Pathology, Brigham and Women's Hospital · Funded by Blavatnik Center for Computational Biomedicine Award

- **Juan Jose Carmona**
  0000-0002-3029-4658 · juancarmona · jcveritas
  Philips Healthcare, Cambridge, MA, USA; Philips Research North America, Cambridge, MA, USA

- **Elana J. Fertig**
  0000-0003-3204-342X · ejfertig
  Department of Oncology, Department of Biomedical Engineering, Department of Applied Mathematics and Statistics, Convergence Institute, Johns Hopkins University · Funded by Lustgarten Foundation; Allegheny Health Network; Emerson Foundation (640183); National Cancer Institute (U01CA212007, U01CA253403, P30CA006973); National Institute of Dental and Cranofacial Research (R01DE027809)

- **Alexandr A. Kalinin**
  0000-0003-4563-3226 · alxndrkalinin · alxndrkalinin
  Medical Big Data Group, Shenzhen Research Institute of Big Data, China; Department of Computational Medicine and Bioinformatics, University of Michigan, USA

- **Beth Signal**
  0000-0002-6839-2392 · betsig
  School of Medicine, College of Health and Medicine, University of Tasmania

- **Benjamin J. Lengerich**
  0000-0001-8690-9554 · blengerich
  Computer Science Department, Carnegie Mellon University

- **Timothy J. Triche, Jr.**
  0000-0001-5665-946X · ttriche
  Center for Epigenetics, Van Andel Research Institute; Department of Pediatrics, College of Human Medicine, Michigan State University; Department of Translational Genomics, Keck School of Medicine, University of Southern California · Funded by National Institute of Allergy and Infectious Diseases (R21AI153997); Michelle Lunn Hope Foundation; Folz Fund for Cancer Research; Grand Rapids Community Foundation

- **Simina M. Boca**
[0000-0002-1400-3398](#) ·  [SiminaB](#)
  Innovation Center for Biomedical Informatics, Georgetown University Medical Center; Department of Oncology, Georgetown University Medical Center; Department of Biostatistics, Bioinformatics and Biomathematics, Georgetown University Medical Center; Cancer Prevention and Control Program, Lombardi Comprehensive Cancer Center · Funded by National Institutes of Health (R21 CA220398)

# Introduction

Machine learning is a modern approach to problem-solving and task automation. In particular, machine learning is concerned with the development and applications of algorithms that can recognize patterns in data and use them for predictive modeling, as opposed to having domain experts developing rules for prediction tasks manually. Artificial neural networks are a particular class of machine learning algorithms and models that evolved into what is now described as "deep learning". Deep learning encompasses neural networks with many layers and the algorithms that make them perform well. These neural networks comprise artificial neurons arranged into layers and are modeled after the human brain, even though the building blocks and learning algorithms may differ [1]. Each layer receives input from previous layers (the first of which represents the input data), and then transmits a transformed version of its own weighted output that serves as input into subsequent layers of the network. Thus, the process of "training" a neural network is the tuning of the layers' weights to minimize a cost or loss function that serves as a surrogate of the prediction error. The loss function is differentiable so that the weights can be automatically updated to attempt to reduce the loss. Deep learning uses artificial neural networks with many layers (hence the term "deep"). Given the computational advances made in the last decade, it can now be applied to massive data sets and in innumerable contexts. In many circumstances, deep learning can learn more complex relationships and make more accurate predictions than other methods. Therefore, deep learning has become its own subfield of machine learning. In the context of biological research, it has been increasingly used to derive novel insights from high-dimensional biological data [2]. For example, deep learning has been used to predict protein-drug binding kinetics [3], to identify the lab-of-origin of synthetic DNA [4], and to uncover the facial phenotypes of genetic disorders [5].

To make the biological applications of deep learning more accessible to scientists who have some experience with machine learning, we solicited input from a community of researchers with varied biological and deep learning interests. These individuals collaboratively contributed to this manuscript's writing using the GitHub version control platform [6] and the Manubot manuscript generation toolset [7]. The goal was to articulate a practical, accessible, and concise set of guidelines and suggestions to follow when using deep learning (Figure 1). For readers who are new to machine learning, we recommend reviewing general machine learning principles [8] before getting started with deep learning.

# Ten Quick Tips for Deep Learning in Biology

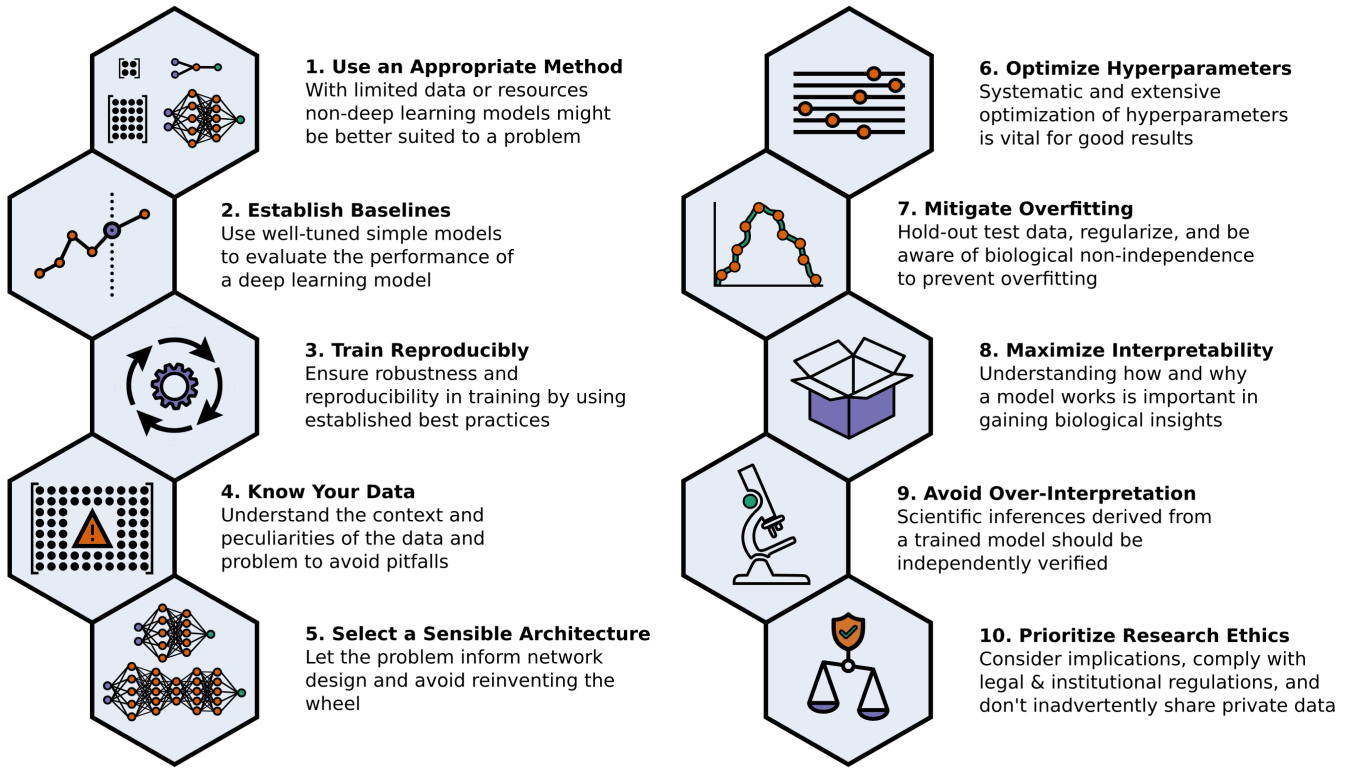

**Figure 1:** A summary overview of the 10 tips for using deep learning in biological research.

In the course of our discussions, several themes became clear: the importance of understanding and applying machine learning fundamentals as a baseline for utilizing deep learning, the necessity for extensive model comparisons with careful evaluation, and the need for critical thought in interpreting results generated by deep learning, among others. The major similarities between deep learning and traditional computational methods also became apparent. Although deep learning is a distinct subfield of machine learning, it is still a subfield. It is subject to the many limitations inherent to machine learning, and most best practices for machine learning [9,10] also apply to deep learning. As with all computational methods, deep learning should be applied in a systematic manner that is reproducible and rigorously tested. Ultimately, the tips we collate range from high-level guidance to best practices for implementation. It is our hope that they will provide actionable, deep learning-specific instruction for both new and experienced deep learning practitioners. By making deep learning more accessible for use in biological research, we aim to improve the overall usage and reporting quality of deep learning in the literature, and to enable increasing numbers of researchers to utilize these state-of-the art techniques effectively and accurately.

# Tip 1: Decide whether deep learning is appropriate for your problem

In recent years, the number of projects and publications implementing deep learning in biology has risen tremendously [11–13]. This trend is likely driven by deep learning's usefulness across a range of scientific questions and data modalities, and can contribute to the appearance of deep learning as a panacea for nearly all modeling problems. Indeed, neural networks are universal function approximators and derive tremendous power from this theoretical capacity to learn any function [14,15]. However, in reality, deep learning is not suited to every modeling situation and can be significantly limited by its large demands for data, computing power, programming skill, and modeling expertise.

While large amounts of high-quality data may be available in the areas of biology where data collection is thoroughly automated, such as DNA sequencing, areas of biology that rely on manual data collection may not possess enough data to train and apply deep learning models effectively. Though there are methods that try to increase the amount of training data, such as data augmentation (in which existing data is slightly manipulated in an attempt to yield "new" samples) and weak supervision (in which simple labeling heuristics are combined to produce noisy, probabilistic labels) [16], these methods cannot overcome substantial data shortages.

In the fields of computer vision and natural language processing, deep neural networks are routinely trained on sample sizes ranging from hundreds of thousands to millions of training examples. Datasets of this size are often not available in many biological contexts. Still, it has been found that, under certain circumstances, deep learning can be considered for datasets with only one hundred samples per class [17]. Nonetheless, deep learning is generally best suited for datasets that contain orders of magnitude more samples.

Training deep learning models often requires extensive computing infrastructure and patience to achieve state-of-the-art performance [18]. In some deep learning contexts, such as generating human-like text, state-of-the-art models have over one hundred billion parameters [19] and require very costly and time-consuming training procedures [20]. These types of large language models are being used in biology to learn representations of protein sequences [21–23]. Even those most deep learning applications in biology rarely require this much training, they can still require computational resources beyond those available on consumer-grade devices such as laptops or office desktops. Specialized hardware such as discrete graphics processing units (GPUs) and custom deep learning accelerators can dramatically reduce the time and cost required to train models [13]. Still, this hardware is not universally accessible, and cloud-based rentals add additional cost and complexity. These specialized hardware solutions are likely to be more broadly available as deep learning becomes more popular. For example, recent-generation iPhones already have such hardware. In contrast to the large scale computational demands of deep learning, traditional machine learning models can often be trained on laptops (or even on a $5 computer [24]) in seconds to minutes. Therefore, due to this enormous disparity in resource demand alone, traditional machine learning approaches may be desirable in various biological applications.

Beyond requiring more data and computational capacity, building and training deep learning models often requires more expertise than training traditional machine learning models. While popular programming frameworks for deep learning such as Tensorflow [25] and PyTorch [26] allow users to create and deploy entirely novel model architectures, this flexibility combined with the rapid development of the deep learning field has resulted in large and complex frameworks that can be daunting to new users. For readers new to software development but experienced in biology, gaining computational skills while also interfacing with such complex industrial-grade tools can be a prohibitive challenge. Conversely, traditional machine learning methods are generally more straightforward to apply and are also more accessible through popular frameworks [27]. Furthermore, there are currently more tools for automating the model selection and training process for traditional machine learning models than for deep learning models. For example, automated machine learning (AutoML) tools, such as TPOT [28] and Turi Create [29], are able to test and optimize multiple machine learning models automatically, and can allow users to achieve competitive performance with only a few lines of code. There are efforts underway to extend these and other automation frameworks to reduce the expertise required to build and use deep learning models. For example, TPOT, Turi Create, and AutoKeras [30] are already capable of abstracting away much of the programming required for "standard" deep learning tasks, and high-level interfaces such as Keras [31] and Fastai [32], make it increasingly straightforward to design and test custom deep learning architectures In the future, projects such as these are likely to make deep learning accessible to a much wider range of researchers.

Despite these limitations, deep learning is strongly indicated over traditional machine learning for specific research questions and problems. In general, these include problems that feature hidden patterns across the data, complex relationships, and interrelated variables. Problems in computer vision and natural language processing often exhibit these very features, which helps explain why these areas were some of the first to experience significant breakthroughs during the recent deep learning revolution [33]. As long as large amounts of accurate and labeled data are available, applications to areas of life sciences with related data characteristics, such as genetic medicine [34], radiology [35], microscopy [36], and pharmacovigilance [37], are similarly likely to benefit from deep learning techniques. For example, Ferreira et al. used deep learning to recognize individual birds from images [38] despite this problem being very difficult historically. By combining automatic data collection using radio-frequency identification tags with data augmentation and transfer learning, the authors were able to use deep learning to achieve 90% accuracy across several species. Another research area where deep learning excels is generative modeling, where new samples are created based on the training data [39]. For example, deep learning can generate realistic compendia of gene expression samples [40]. One other area of machine learning that has been revolutionized by deep learning is reinforcement learning, which is concerned with training agents to interact with an environment [41]. Reinforcement learning has been applied to design druglike small molecules [42]. Overall, initial evaluation as to whether similar problems (including analogous ones in other domains) have been solved successfully using deep learning can inform researchers about the potential for deep learning to address their needs.

On the other hand, depending on the amount and type of data available and the nature of the problem set, deep learning may not always outperform conventional methods. As an illustration, Rajkomar et al. [43] found that simpler baseline models achieved performance comparable with deep learning in several clinical prediction tasks using electronic health records. Another example is provided by Koutsoukas et al., who benchmarked several traditional machine learning approaches against deep neural networks for modeling bioactivity data on moderately sized datasets [44]. The researchers found that while well-tuned deep learning approaches generally tend to outperform conventional classifiers, simple methods such as Naive Bayes classification tend to outperform deep learning as the dataset's noise increases. Similarly, Chen et al. [45] tested deep learning and a variety of traditional machine learning methods such as logistic regression and random forests on five different clinical datasets. They found that traditional methods matched or exceeded the accuracy of the deep learning model in all cases despite requiring an order of magnitude less training time.

Therefore, in conclusion, deep learning should only be used after a robust consideration of its strengths and weaknesses for the problem at hand. After choosing deep learning as a potential solution, practitioners should still consider traditional methods as performance baselines and use the scientific method to compare the performance of deep learning to that of traditional methods.

## Tip 2: Use traditional methods to establish performance baselines

Deep learning requires practitioners to consider a larger number and variety of tuning parameters (that is, algorithmic settings) than more traditional machine learning methods. These settings are often called hyperparameters. Their extensiveness can make it easy to fall into the trap of performing an unnecessarily convoluted analysis. Hence, before applying deep learning to a given problem, it is ideal to implement simpler models with fewer hyperparameters at the beginning of each study. Such models include logistic regression, random forests, k-nearest neighbors, Naive Bayes, and support vector machines. They can help to establish baseline performance expectations and the difficultly of a specific prediction problem. While performance baselines available from existing literature can also serve as helpful guides, an implementation of a simpler model that uses the same software framework as planned for deep learning can greatly help with assessing the correctness of data processing steps, performance evaluation pipelines, resource requirement estimates, and computational performance estimates. Furthermore, in some cases, it can even be useful to combine simpler baseline models with deep neural networks, as such hybrid models can improve generalization performance, model interpretability, and confidence estimation [46,47].

Another potential pitfall arises from comparing the performance of baseline conventional models trained with default settings with the performance of deep learning models that have undergone rigorous tuning and optimization. Since conventional off-the-shelf machine learning algorithms (for example, support vector machines and random forests) are also likely to benefit from hyperparameter tuning, such incongruity prevents the comparison of equally optimized models and can lead to false conclusions about model efficacy. Hu and Greene [48] discuss this under the umbrella of what they call the "Continental Breakfast Included" effect. They describe how the unequal tuning of hyperparameters across different learning algorithms can especially skew evaluation when the performance of an algorithm varies substantially with modest changes to its hyperparameters. Therefore, practitioners should tune the settings of both traditional machine learning and deep learning-based methods before making claims about relative performance differences. Performance comparisons among machine learning and deep learning models are only informative when the models are equally well optimized.

In sum, practitioners are encouraged to create and fully tune several traditional models and standard pipelines before implementing a deep learning model.

## Tip 3: Understand the complexities of training deep neural networks

Correctly training deep neural networks is non-trivial. There are many different options and potential pitfalls at every stage. To get good results, one must often train networks across a wide range of different hyperparameter settings. Such training can be made more difficult by the demanding nature of these deep networks, which often require extensive time investments into tuning and computing infrastructure to achieve state-of-the-art performance [18]. Furthermore, this experimentation is often noisy, which necessitates increased repetition and exacerbates the challenges inherent to deep learning. On the whole, all code, random seeds, parameters, and results must be carefully corralled using general coding standards and best practices (for example, version control [49] and continuous integration [50]) to be reproducible and interpretable [51,52]. For application-based research, this organization is also fundamental to the efficient sharing of research work and the ability to keep models up to date as new data becomes available.

One specific reproducibility pitfall that is often missed in applying deep learning is the default use of non-deterministic algorithms by CUDA/CuDNN backends when using GPUs. That is, the CUDA/CuDNN architectures that facilitate the parallelized computing that power state-of-the-art deep learning often use algorithms by default that produce different outcomes from iteration to iteration. Therefore, achieving reproducibility in this context requires explicitly specifying the use of deterministic algorithms (which are typically available within deep learning libraries), which is distinct from the setting of random seeds that typically achieve reproducibility by controlling pseudorandom deterministic procedures such as shuffling and initialization [53].

Similar to the suggestions above about starting with simpler models, starting with relatively small networks and then increasing the size and complexity as needed can help prevent practitioners from wasting significant time and resources on running highly complex models that feature numerous unresolved problems. Again, practitioners must beware of the choices made implicitly (that is, by default settings) by deep learning libraries. These seemingly trivial details, such as the selection of optimization algorithm, can have significant effects on model performance. For example, adaptive methods often lead to faster convergence during training but may lead to worse generalization performance on independent datasets [54]. These nuanced elements are easy to overlook, but it is critical to consider them carefully and to evaluate their potential impact.

In short, researchers should use smaller and simpler networks to enable faster prototyping, follow general software development best practices to maximize reproducibility, and check software documentation to understand default choices.

## Tip 4: Know your data and your question

Having a well defined scientific question and a clear analysis plan is crucial for carrying out a successful deep learning project. Just like it would be inadvisable to set foot in a laboratory and begin experiments without having a defined endpoint, a deep learning project should not be undertaken without defined goals. Foremost, it is important to assess if a dataset exists that can answer the biological question of interest using a deep learning-based approach. If so, obtaining this data (and associated metadata) and reviewing the study protocol should be pursued as early on in the project as possible. This can help to ensure that data is as expected and can prevent the wasted time and effort that occur when issues are discovered later on in the analytic process. For example, a publication or resource might purportedly offer an appropriate dataset that is found to be inadequate upon acquisition. The data may be unstructured when it is supposed to be structured, crucial metadata such as sample stratification might be missing, or the usable sample size may be different than expected. Any of these data issues might limit a researcher's ability to use deep learning to address the biological question at hand or might otherwise require adjustment before it can be used. Data collection should also be carefully documented, or a data collection protocol should be created and specified in the project documentation.

Information about the resources used, download dates, and dataset versions are critical to preserve. Doing so will help to minimize operational confusion and will increase the reproducibility of the analysis. Best practices for reproducibility also include sharing the collected dataset and metadata upon publication of the study, ideally in a public dataset repository if there are no ethical or privacy concerns and no copyright restrictions. While recommended and recognized dataset repositories may differ across disciplines, a list of general dataset repositories includes the Dryad repository [55] (https://datadryad.org/), Figshare [56] (https://figshare.com), Zenodo [57] (https://zenodo.org), and the Open Science Framework [58] (https://osf.io). In addition, Gundersen et al. [59] provide useful checklists summarizing best data sharing practices for reproducible research and open science.

Once the dataset is obtained, it is important to learn why and how the data were collected before beginning analysis. The standardized metadata that exists in many fields can help with this (for example, see [60]). If at all possible, consulting with a subject matter expert who has experience with the type of data being used will minimize guesswork and likely increase the success rate of a deep learning project. For example, one might presume that data collected to test the impact of an intervention are derived from a randomized controlled trial. However, this is not always the case, as ethical or practical concerns often necessitate an observational study design that features prospectively or retrospectively collected data. In order to ensure similar distributions of important characteristics across study groups in the absence of randomization, such a study may have selected individuals in a fashion that best matches attributes such as age, gender, or weight. Passively collected datasets can have their own peculiarities, and other study designs can include samples that originate from the same study site, the oversampling of ethnic groups or zip codes, or sample processing differences. Such information is critical to accurate data analysis, and so it is imperative that practitioners learn about study design assumptions and data specificities prior to performing modeling. Other study design considerations that should not be overlooked include knowing whether a study involves biological or technical replicates or both. For example, the existence in a dataset of samples collected from the same individuals at different time points can have significant effects on analyses that make assumptions about sample size and independence (that is, non-independence can lower the effective sample size). Another potential issue is the existence of systematic biases, which can be induced by confounding variables and can lead to artifacts or so-called "batch effects." Consequently, models may learn to rely on the correlations that these systematic biases underpin, even though they are irrelevant to the scientific context of the study. This can lead to misguided predictions and misleading conclusions [61]. Unsupervised learning and other exploratory analyses can help identify such biases in these datasets before applying a deep learning model.

Overall, practitioners should thoroughly study their data and understand its context and peculiarities *before* moving on to performing deep learning.

## Tip 5: Choose an appropriate data representation and neural network architecture

Neural network architecture refers to the number and types of layers in the network and how they are connected. While certain best practices have been established by the research community [62], architecture design choices remain largely problem-specific and are vastly empirical efforts requiring extensive experimentation. Furthermore, as deep learning is a quickly evolving field, many recommendations are often short-lived and are frequently replaced by newer insights supported by recent empirical results. This is further complicated by the fact that many recommendations do not generalize well across different problems and datasets. Therefore, choosing how to represent data and design an architecture is closer to an art than a science. That said, there are some general principles to follow when experimenting.

First and foremost, knowledge of the available data and scientific question should inform data representation and architectural design choices. For example, if the dataset is an array of measurements with no natural ordering of inputs (such as gene expression data), multilayer perceptrons may be effective. These are the most basic type of neural network. They are able to learn complex non-linear relationships across the input data despite their relative simplicity. Similarly, if the dataset is comprised of images, convolutional neural networks (CNNs) are a good choice because they emphasize local structures and adjacency within the data. CNNs may also be a good choice for learning on sequences. Recent empirical evidence suggests that they can outperform canonical sequence learning techniques such as recurrent neural networks and the closely related long short-term memory networks in some cases [63]. Transformers are powerful sequence models [64] but require extensive data and computing power to train from scratch. Accessible high-level overviews of different neural network architectures are provided in [65] and [66].

Deep learning models typically benefit from increasing the amount of labeled training data. Large amounts of data help to avoid overfitting and increase the likelihood of achieving top performance on a given task. If there is not enough data available to train a well-performing model, transfer learning should be considered. In transfer learning, a model whose weights were generated by training on another dataset is used as the starting point for training [67]. Transfer learning is most useful when the pre-training and target datasets are of similar nature [67]. For this reason, it is important to search for similar datasets that are already available. However, even when similar datasets do not exist, transferring features can still improve model performance compared with random feature initialization. For example, Rajkomar et al. showed advantages of ImageNet-pretraining [68] for a model that is applied to grayscale medical image classification [69]. Pre-trained models can be obtained from public repositories, such as Kipoi [70] for genomics or Hugging Face [71] for biomedical text [72], protein sequences [22], and chemicals [73]. Moreover, learned features could be helpful even when a pre-training task is different from a target task [74].

Recently, the concept of self-supervised learning, which is closely related to pre-training and transfer learning, has seen an increase in popularity [75]. Self-supervised learning leverages large amounts of unlabeled data and uses naturally available information as labels for supervised learning. Thus, self-supervised learning is sometimes also described as autonomous supervised learning. Using self-supervised learning, a model can be pre-trained on a related task before it is trained on the target task. Another related approach is multi-task learning, which simultaneously trains a network for multiple separate tasks that share features. In fact, multi-task learning can be used separately or even in combination with transfer learning [76].

This tip can be distilled into two main action points for practitioners: first, they should base the network's architecture on knowledge of the problem and, second, they should take advantage of similar existing data or pre-trained deep learning models.

## Tip 6: Tune your hyperparameters extensively and systematically

Given at least one hidden layer, a non-linear activation function, and a large number of hidden units, multi-layer neural networks can approximate arbitrary continuous functions that relate input and output variables [15,77]. Deeper architectures that feature additional hidden layers and an increasing number of overall hidden units and learnable weight parameters (the so-called increasing "capacity" of neural networks) allow for solving increasingly complex problems. However, this increased capacity results in many more parameters to fit and hyperparameters to tune, which can pose additional challenges during model training. In general, one should expect to systematically evaluate the impact of numerous hyperparameters when applying deep neural networks to new data or challenges. Hyperparameters typically manifest as choices of optimization algorithms, loss function, learning rate, activation functions, number of hidden layers and hidden units, size of the training batches, and weight initialization schemes. Moreover, additional hyperparameters are introduced by common techniques that facilitate training deeper architectures. These include regularization penalties, dropout [78], and batch normalization [79], which can reduce the effect of the so-called vanishing or exploding gradient problem when working with deep neural networks.

This wide array of potential parameters can make it difficult to evaluate the extent to which neural network methods are well suited to solving a task. For example, it can be unclear to practitioners whether previously successful deep learning applications were the result of general model suitability for the data at hand or interactions between unique data attributes and specific hyperparameter settings. As a result, a lack of clarity about how extensive arrays of hyperparameters were tested or chosen can hamper method developers as they attempt to compare techniques. This effect also has implications for those seeking to use existing deep learning methods, as performance estimates from deep neural networks are often provided after tuning. The implication is that attaining performance numbers that match those reported in publications is likely to require significant effort towards temporally expensive hyperparameter optimization. Strategies for tuning hyperparameters include exhaustive grid search, random search, or Bayesian optimization and other specialized techniques. Tools such as Keras Tuner (https://keras-team.github.io/keras-tuner/) and Ray Tune (https://docs.ray.io/en/latest/tune/index.html) support best practices for hyperparmeter optimization.

To get the best performance from your model, researchers should be sure to systematically optimize their hyperparameters on the training dataset and report both the selected hyperparameters and the hyperparameter optimization strategy.

## Tip 7: Address deep neural networks' increased tendency to overfit the dataset

Overfitting is a challenge inherent to machine learning in general and is one of the most significant challenges you'll face when applying deep learning specifically. Overfitting occurs when a model fits patterns in the training data so closely that it includes non-generalizable noise or non-scientifically relevant perturbations in the relationships it learns. In other words, the model fits patterns that are overly specific to the data it is training on rather than learning general relationships that hold across similar datasets. This subtle distinction is made clearer by seeing what happens when a model is tested on data to which it was not exposed during training: just as a student who memorizes exam materials struggles to correctly answer questions for which they have not studied, a machine learning model that has overfit to its training data will perform poorly on unseen test data. Deep learning models are particularly susceptible to overfitting due to their relatively large number of parameters and associated representational capacity. Just as some students may have greater potential for memorization, deep learning models seem more prone to overfitting than machine learning models with fewer parameters. However, having a large number of parameters does not always imply that a neural network will overfit [80].

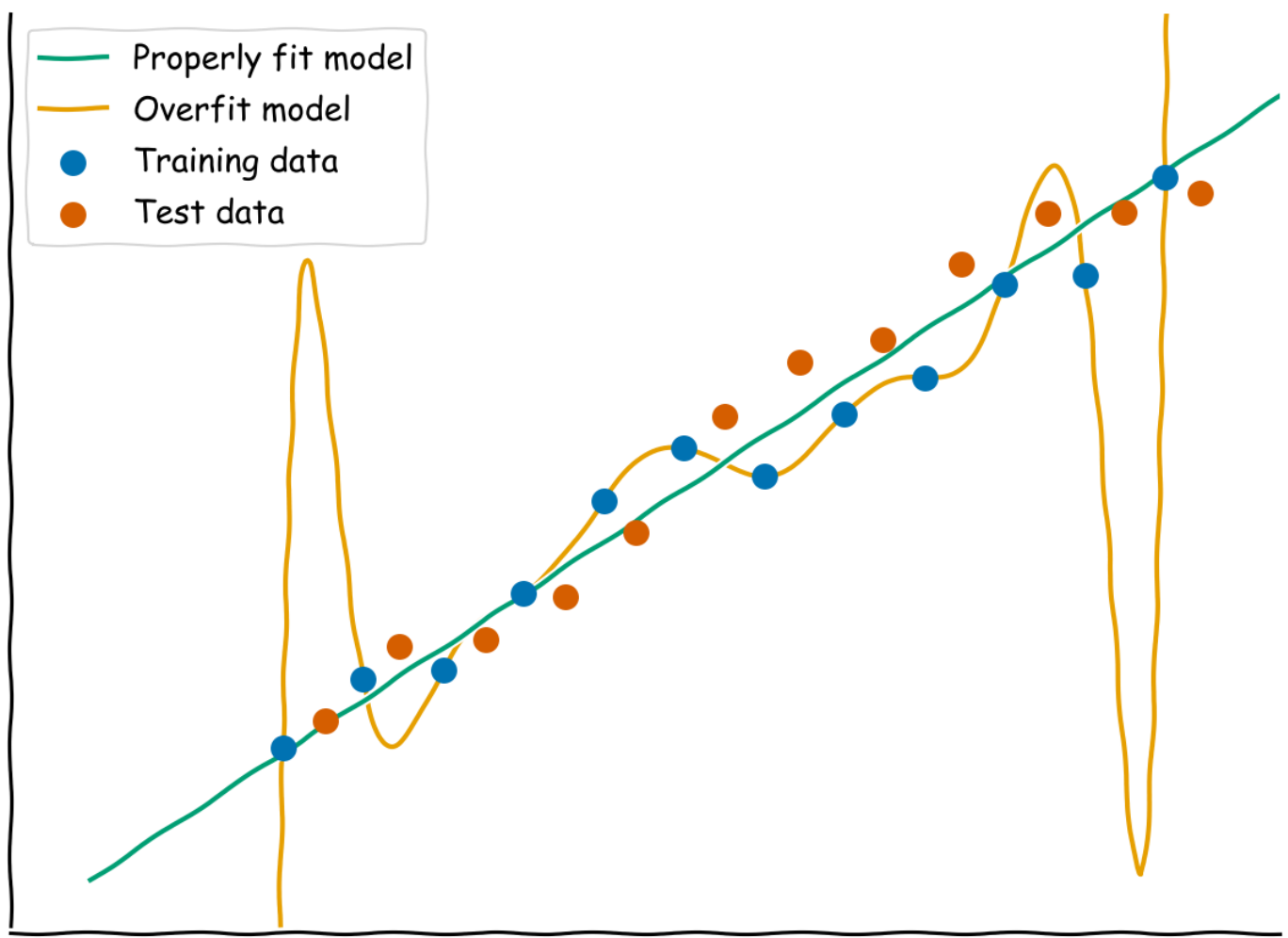

**Figure 2:** A visual example of overfitting and failure to generalize. While a high-degree polynomial achieves high accuracy on its training data, it performs poorly on the test data that have not been seen before. That is, the model has memorized the training dataset specifically rather than learning a generalizable pattern that represents data of this type. In contrast, a simple linear regression works equally well on both datasets.

In general, one of the most effective ways to combat overfitting is to detect it in the first place. One way to do this is to split the main dataset being worked on into three independent parts: a training set, a tuning set (also commonly called a validation set in the machine learning literature), and a test set. These three partitions allow you to optimize models by iterating between model learning on the training set and hyperparameter evaluation on the tuning set without affecting the final model assessment on the test set. A researcher can compare the model's performance on the training and tuning data to assess how overfit (i.e. non-generalizable) the model is. The data used for testing should be "locked away" and used only once to evaluate the final model after all training and tuning steps are completed. This type of approach is necessary for evaluating the generalizability of models without the biases that can arise from learning and testing on the same data [81,82]. While a slight drop in performance from the training set to the test set is normal, a significant drop is a clear sign of overfitting. See Figure 2 for a visual demonstration of an overfit model that performs poorly on test data.

There are a variety of techniques to reduce overfitting, including data augmentation and regularization techniques [83,84]. Another way to reduce overfitting, as described by Chuang and Keiser, is to identify the baseline level of memorization that is occurring by training on data that has its labels randomly shuffled [85]. By comparing the model performance with the shuffled data to that achieved with the actual data [85], a practitioner can identify overfitting as a model that performs no better on real data. This suggests that any predictive capacity is not due to data-driven signal. One important caveat when working with partitioned data is the need to apply transformation and normalization procedures equally to all datasets. The parameters required for such procedures (for example, quantile normalization, a common standardization method when analyzing gene-expression data) should only be derived from the training data, and not from the tuning or test data. Additionally, many conventional metrics for classification (e.g. area under the receiver operating characteristic curve or AUROC) have limited utility in cases of extreme class imbalance [86]. Therefore, model performance should be evaluated with a carefully picked panel of relevant metrics that make minimal assumptions about the composition of the testing data [87].

When working with biological and medical data, one must also carefully consider potential sources of bias and/or non-independence when defining training and test sets. For example, a deep learning model for pneumonia detection in chest X-rays appeared to performed well within the hospitals providing the training data, but then failed to generalize to other hospitals [88]. This resulted from the deep learning model picking up on signal related to which hospital the images were from and represents a type of artifact or "batch effect" that practitioners must be vigilant towards. When dealing with sequence data, holding out test data that are evolutionarily related or that share structural homology to the training data can result in overfitting that is hard to detect due to the inherent relatedness of the partitioned data [89]. In such situations, simply holding out test data selected from a random partition of the training data can be insufficient. The best remedy for identifying confounding variables is to know your data and to test models on truly independent data.

In essence, practitioners should split data into training, tuning, and single-use testing sets to assess the performance of the model on data that can provide a reliable estimate of its generalization performance. Furthermore, practitioners should be cognizant of the danger of skewed or biased data artificially inflating performance.

## Tip 8: Deep learning models can be made more transparent

While model interpretability is a broad concept, in much of the machine learning literature it refers to the ability to identify the discriminative features that influence or sway the predictions. In certain cases, the goal behind interpretation is to understand the underlying data generating processes and biological mechanisms [90]. In other cases, the goal is to understand why a model made the prediction that it did for a specific example or set of examples. Machine learning models vary widely in terms of interpretability: some are fully transparent while others are considered "black-boxes" that make predictions with little ability to examine why. Logistic regression and decision tree models are generally considered interpretable [91]. In contrast, deep neural networks are often considered among the most difficult to interpret naively because they can have many parameters and non-linear relationships.

Knowing which of the input variables influences a model's predictions, and potentially in what ways, can help with the application or extrapolation of machine learning models. This is particularly important in biomedicine. Subsequent decision making often requires human input, and models are employed with the hope of better understanding why relationships exist in the first place. Furthermore, while prediction rules can be derived from high-throughput molecular datasets, most affordable clinical tests still rely on lower-dimensional measurements of a limited number of biomarkers. Therefore, it is often unclear how to translate the predictive capacity of deep learning models that encompass non-linear relationships between countless input variables into clinically digestible terms. As a result, selecting which biomarkers to use for decision making remains an important modeling and interpretation challenge. In fact, many authors attribute a lower uptake of deep learning tools in healthcare to interpretability challenges [92,93]. Nonetheless, strategies to interpret both machine learning and deep learning models are rapidly emerging, and the literature on the topic is growing exponentially [94]. Instead of recommending specific methods for either deep learning-specific or general-purpose model interpretation, we suggest consulting a freely available and continually updated textbook [95].

## Tip 9: Don't over-interpret predictions

After training an accurate deep learning model, it is natural to want to use it to deduce relationships and inform scientific findings. However, be careful to interpret the model's predictions correctly. Given that deep learning models can be difficult to interpret intuitively, there is often a temptation to over-interpret the predictions in indulgent or inaccurate ways. In accordance with the classic statistical saying "correlation doesn't imply causation," predictions by deep learning models rarely provide causal relationships. Accurately predicting an outcome does not mean that a causal mechanism has been learned, even when predictions are extremely accurate. In a poignant example, authors evaluated the capacities of several models to predict the probability of death for patients with pneumonia admitted to an intensive care unit [96,97]. The neural network model achieved the best predictive accuracy. However, after fitting a rule-based model to understand the relationships inherent to their data better, the authors discovered that the hospital data implied the rule "$\mathrm{HasAsthma}(x) \Rightarrow \mathrm{LowerRisk}(x)$." This rule contradicts medical understanding, as having asthma does not make pneumonia better! Nonetheless, the data supported this rule, as pneumonia patients with a history of asthma tended to receive more aggressive care. The neural network had, therefore, also learned to make predictions according to this rule despite the fact that it has nothing to do with causality or mechanism. Therefore, it would have been disastrous to guide treatment decisions according to the predictions of the neural network, even though the neural network had high predictive accuracy.

It is critical to avoid over-interpreting deep learning models by viewing them for what they are: complex statistical models trained on high dimensional data. If causal inference is desired, special techniques for causal inference are required [98].

## Tip 10: Actively consider the ethical implications of your work

While deep learning continues to be a powerful, transformative tool within life sciences research—spanning from basic biology and pre-clinical science to varied translational approaches and clinical studies—it is important to comment on ethical considerations. For instance, despite the fact that deep learning methods are helping to increase medical efficiency through improved diagnostic capability and risk assessment, certain biases may be inadvertently introduced into models related to patient age, race, and gender [99]. Deep learning practitioners may make use of datasets not representative of diverse populations and patient characteristics [100], thereby contributing to these problems.

Therefore, it is important to think thoroughly and cautiously about deep learning applications and their potential impact to persons and society. This includes being mindful of possible harms, injustices, and other types of wrongdoing. At a minimum, practitioners must ensure that, wherever relevant, their life sciences projects are fully compliant with local research governance and approval policies, legal requirements, Institutional Review Board policies, and any other relevant bodies and standards. Moreover, we offer below three tangible, action-oriented recommendations to further empower and enrich deep learning researchers.

First, just as it is a best practice to keep a project-specific or programming-related issue tracker detailing known bugs and other technical issues, practitioners should get into the habit of keeping an active *ethics register*. In this register, ethical concerns can be raised, recorded, and resolved, exactly as software problems are triaged and fixed. Because projects using deep learning usually rely on writing code, an ethics register can be a part of the issue tracker in the version control system for the software itself. By colocating the two, practitioners can operationalize the concept that ethical problems are "bugs" that must be resolved, not nice-to-haves that can be considered at some indefinite point in the future. For practitioners intending to distribute trained models, having an ethics register can also facilitate creating a *model card* [101], a short document specifying the domains in which the model's performance was validated (for example, which model organism was used), how the performance was benchmarked, and known limitations and concerns. Second, to help foster a conscious ethics-oriented mindset, researchers should consider expanding journal clubs to include scholarly and popular articles detailing real-world ethics issues relevant to their scientific fields. This will help researchers to think more holistically and judiciously about their work and its implications. Third, we encourage individual- and team-level participation in professional societies [102] and other types of organizations [103] and events [104] related to the domains of AI and data ethics as well as bioethics. This will encourage a sense of community and intellectual engagement, and will keep practitioners abreast of cutting-edge insights and emerging professional standards.

Furthermore, practitioners may encounter datasets that cannot be shared, such as ones for which there would be significant ethical or legal issues associated with their release [105]. Examples of such data include classified or confidential data, biological data related to trade secrets, medical records, or other personally identifiable information [106]. While deep learning models can capture information-rich abstractions of multiple features of the data during the training process, these features may be more prone to leak the data that they were trained over if the model is shared or allowed to be queried with arbitrary inputs [107,108]. In other words, the complex relationships learned about the input data can potentially be used to infer characteristics about the original dataset, which reflects the facts that the strengths that imbue deep learning with its great predictive capacity may also raise the level of risk surrounding data privacy. Therefore, while there is tremendous promise for deep learning techniques to extract information that cannot readily be captured by traditional methods [109], it is imperative not to share models trained on sensitive data. This also holds true for certain traditional machine learning methods that learn by capturing specific details of the full training data (for example, *k*-nearest neighbors models).

Techniques to train deep neural networks without sharing unencrypted access to data are being advanced through implementations of homomorphic encryption, which serves to enable equivalent prediction on data that is encrypted end-to-end [110,111]. Privacy-preserving techniques [112], such as differential privacy [113–115], can help to mitigate risks as long as the assumptions underlying these techniques are met. While these methods provide a path towards a future where trained models and their predictions can be shared, more software development and theoretical advances will be required to make these techniques easy to apply correctly in many settings. Unless using these techniques, researchers must not share the weights or provide arbitrary access to the predictions of models trained on sensitive data.

# Conclusion

Collectively, we have presented practical tips that emphasize cutting-edge insights and describe evolving professional standards. In addition, a number of our points are focused on safeguarding against the risks inherent to data science and deep learning. These risks include the over- and mis-interpretation of models, poor generalizability, and the potential to harm others. However, we want to strongly emphasize that when used ethically and responsibly, deep learning techniques have the potential to add tremendous value across a diverse range of contexts. After all, these techniques have already shown a remarkable capacity to meet or exceed the performance of both humans and traditional algorithms and have the potential to uncover biomedical insights that drive discovery and innovation. By taking a comprehensive and careful approach to deep learning based on critical thinking about research questions, planning to maintain rigor, and discerning how work might have far-reaching consequences with ethical dimensions, the life science community can advance reproducible, interpretable, and high-quality science that is enriching and beneficial for both scientists and society.

# Competing interests

| Author | Competing interests |
| --- | --- |
| Anthony Gitter | Filed a patent application with the Wisconsin Alumni Research Foundation related to classifying activated T cells. |
| Kun-Hsing Yu | Inventor of a quantitative pathology analytical system (U.S. Patent 10,832,406). This invention is not related to this work. |
| Elana J. Fertig | Scientific Advisory Board member at Viosera Therapeutics. |
| Alexandr A. Kalinin | Co-inventor on 4 patent applications related to machine learning applications in biology. |

All other authors have declared no competing interests.

# Acknowledgements

The authors would like the thank Daniel Himmelstein and the developers of Manubot for creating the software that enabled the collaborative composition of this manuscript. We would also like to thank Fábio Madeira ([0000-0001-8728-9449](#)), Victor Greiff ([0000-0003-2622-5032](#)), Shyam Saladi ([0000-0001-9701-3059](#)), Anshul Kundaje ([0000-0003-3084-2287](#)), Brett K. Beaulieu-Jones ([0000-0002-6700-1468](#)), Paul Brodersen ([0000-0001-5216-7863](#)), Michael M. Hoffman ([0000-0002-4517-1562](#)), and Isaac Lazzeri for their contributions to the discussions that comprised the initial stage of the drafting process. This work has been supported in part by the Biostatistics and Bioinformatics Shared Resource at the H. Lee Moffitt Cancer Center & Research Institute, an NCI designated Comprehensive Cancer Center (P30-CA076292).